\documentstyle[preprint,aps,epsfig]{revtex}

\begin{document}

\title{{\LARGE Electronic structure of nuclear-spin-polarization-induced quantum dots}}
\author{\-{\it Yu. V. Pershin}} \bigskip
\address{
{\small Center for Quantum Device Technology,}\\ {\small
Department of Physics, Clarkson University, Potsdam, NY
13699-5820, USA}\\ \bigskip {\small Grenoble High Magnetic Fields
Laboratory, }\\{\small Max-Planck-Institute f\"{u}r
Festkorperforschung and CNRS,\ \ }\\ {\small BP 166, F-38042
Grenoble Cedex 9, France}}

\maketitle

\bigskip

\begin{abstract}
We study a system in which electrons in a two-dimensional electron
gas are confined by a nonhomogeneous nuclear spin polarization.
The system consists of a heterostructure that has non-zero nuclei
spins. We show that in this system electrons can be confined into
a dot region through a local nuclear spin polarization. The
nuclear-spin-polarization-induced quantum dot has interesting
properties indicating that electron energy levels are
time-dependent because of the nuclear spin relaxation and
diffusion processes. Electron confining potential is a solution of
diffusion equation with relaxation. Experimental investigations of
the time-dependence of electron energy levels will result in more
information about nuclear spin interactions in solids.

\end{abstract}

\bigskip

\section{INTRODUCTION}

The theoretical and experimental researches of quantum dots have
attracted much attention in recent years \cite{Jacak}. Quantum
dots are usually fabricated experimentally by applying
lithographic and etching techniques to impose a lateral structure
onto an otherwise two-dimensional electron system. Lateral
structures introduce electrostatic potentials in the plane of the
two-dimensional electron gas, which confines the electrons to a
dot region. The energy levels of electrons in such quantum dots
are fully quantized like in an atom. In such electrically confined
quantum dots the confining potential can be well represented by a
parabolic potential.

Another method of low-dimensional structure fabrication consists
of the application of spatially inhomogeneous magnetic fields.
There has been proposed several alternative magnetic structures
subsequently realized experimentally. Among them: magnetic dots
using a scanning tunneling microscope lithographic technique
\cite{McCord}, magnetic superlattices by the patterning of
ferromagnetic materials integrated by semiconductors
\cite{Leadbeater91}, type-II superconducducting materials
deposited on conventional heterostructures \cite{Bending}, and
nonplanar two-dimensional electron gas (2DEG) systems grown by a
molecular beam epitaxy \cite{Leadbeater95}. Such systems were
studied theoretically in a series of papers by different authors
\cite{Peeters1,Peeters2,Peeters3,Peeters4,Solimany,Kim1,Kim2,Kim3,Kim4}.

In the present paper we study a quantum dot system which is
different from the quantum dot systems discussed above: (1) the
electrons are confined through local nuclear spin polarization,
(2) the confinement potential is inherently nonparabolic and
time-dependent, it is a solution of the diffusion equation when
considering relaxation, and (3) the dot contains electrons with
only one spin direction. Such system was proposed for the first
time in Ref. \cite{Fleurov}. However, the properties of
Nuclear-Spin-Polarization-Induced Quantum Dots (NSPIQD) have not
been considered thus far and this is the motivation behind the
present investigation. In our calculations we use some ideas from
\cite{Pershin}, where a nuclear-spin-polarization-induced quantum
wire was proposed and investigated.

Electron and nuclear spins interact via the contact hyperfine
interaction. Once the nuclear spins are polarized, the charge
carrier spins feel the effective hyperfine field, ${\bf B}_{hf}$,
which lifts the spin degeneracy. The maximum nuclear field in GaAs
can be as high as $B_{hf}=$5.3T in the limit that all nuclear
spins are fully polarized \cite{Paget}. This high level of nuclear
spin polarization has been achieved experimentally. For example,
the optical pumping of nuclear spins in 2DEG has demonstrated
nuclear spin polarization on the order of $90\%$, \cite{Salis}. A
similarly high polarization has been created by quantum hall edge
states ($85\%$) \cite{Dixon}. The spin splitting due to such a
hyperfine magnetic field is comparable to the Fermi energy of
2DEG. It is important to note that the hyperfine field does not
manifest itself magnetically due to the smallness of the nuclear
magnetic moments. The electrons in the region where nuclear spins
are polarized will preferably occupy the energetically more
favorable states with the spins opposite to ${\bf B}_{hf}$.
Furthermore, the nuclear polarization acts on the electrons as the
effective confining potential. This effective confining potential
can be used to create different nanostructures with polarized
electrons in them. In this paper we consider a
nuclear-spin-polarization-induced quantum dot (NSPIQD).

\begin{figure}[t]
 \centering
  \includegraphics[width=8cm, angle=270]{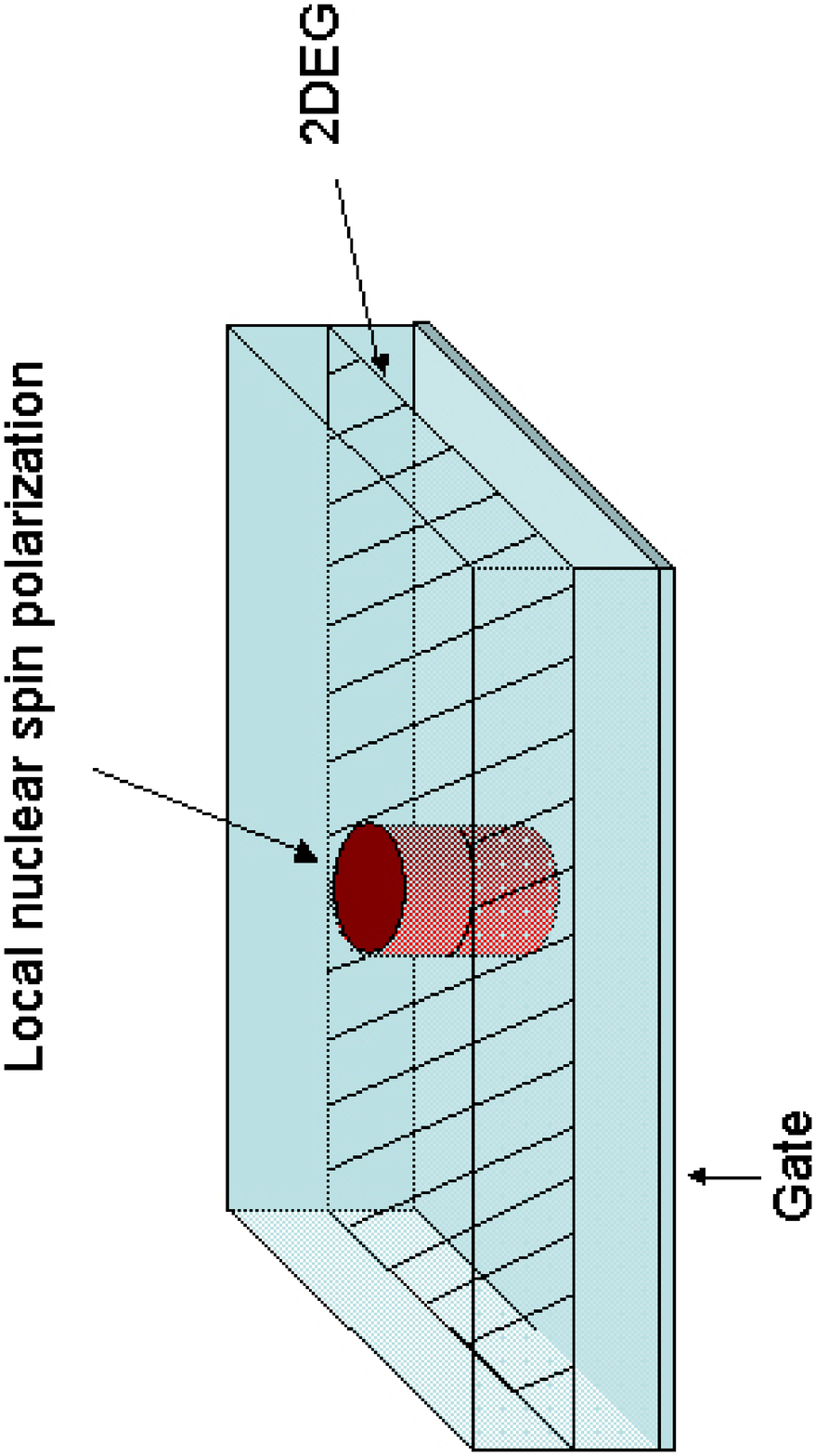}
 \caption{The geometry of the proposed experiment: the NSPIQD is created
 in the region of intersection of the 2DEG with the local nuclear
 spin polarization.}\label{fig1}
\end{figure}

The proposed system is depicted in Fig. \ref{fig1}. The nuclear
spins are polarized homogeneously along the z-axis perpendicular
to the 2DEG in heterostructure by any other suitable experimental
method. For example, the optical nuclear spin polarization \cite
{Lampel,Zahar,Barrett} or the transport polarization
\cite{Kane,Wald} can be used. The region where the nuclear spins
are polarized is indicated by the cylinder in Fig. \ref{fig1}. The
NSPIQD is created in the region of intersection of the 2DEG with
the region of local nuclear spin polarization. The gate electrode
below the 2DEG is used to control the number of electrons in the
NSPIQD. Moreover, the system is subjected to an external magnetic
field along the z-axis. The magnetic field plays an important role
in the nuclear spin polarization process and, under specific
conditions, increases the nuclear spin relaxation time. Assuming a
small magnetic field, we can neglect it in our calculations,
focusing on the effects caused by the confining hyperfine field.
Our paper is organized as follows. In Sec. \ref{sec2} we discuss
the properties of nonhomogeneous nuclear spin polarization and
calculate the evolution of initially-created hyperfine-field
profile which is taken, for simplicity, in the Gaussian form. Time
dependence of the electron states in NSPIQD is studied in Sec.
\ref{sec3}. The conclusions of this investigation are presented in
Sec. \ref{sec4}.

\section{HYPERFINE-FIELD PROFILE } \label{sec2}

Let us assume that the method of optical nuclear spin polarization
is used \cite{Lampel,Zahar,Barrett} to create a NSPIQD. To pattern
a nanostructure it is proposed to illuminate the sample locally
by, for example, putting a mask on it. The usual optical technique
allows one to create the light beams of the width of the order of
the wavelength ($\sim 500nm$). By using near fields optics the
beam width can be sufficiently reduced ($\sim 100nm$). Hence a $1
\mu m$-size NSPIQD can be easily created by the modern
experimental technique.

There are two main mechanisms leading to the time dependence of
the hyperfine field: the nuclear spin relaxation and the nuclear
spin diffusion. We assume that the initial nuclear spin
polarization is homogeneous in the $z$-direction. Then the
hyperfine field evolution is described by the two-dimensional
diffusion equation:

\begin{equation}
\frac{\partial B_{hf}}{\partial t}=D\Delta
B_{hf}-\frac{1}{T_{1}}B_{hf}\text{,}  \label{diffusion}
\end{equation}
accounting for the relaxation processes. Here $D$ is the
spin-diffusion coefficient, $\Delta$ is two-dimensional Laplace
operator, and $T_{1}$ is the nuclear spin relaxation time
\cite{Wolf,Slichter}. The formal solution of Eq.(\ref{diffusion})
can be written as

\begin{equation}
B_{hf}=e^{-\frac{t}{T_1}}\int G\left( {\mathbf r}-{\mathbf r'},t
\right) B_{hf} \left({\mathbf r'},t=0 \right)
d\mathbf{r'}\label{difsol}.
\end{equation}
Here $G\left( \mathbf{r},t \right)=\frac{e^{-%
\frac{\left( \mathbf{r}-\mathbf{r'}\right) ^{2}}{4Dt}}}{4\pi Dt}$
is the Green function of the diffusion equation and $B_{hf}
\left({\mathbf r'},t=0 \right)$ is the initial hyperfine field
profile.

In this paper we consider NSPIQD having the cylindrical symmetry;
that is, the hyperfine field $B_{hf}$ is a function of $r$. In the
simplest case, we can assume the initial condition to be of the
Gaussian form: $B_{hf}\left( r,0\right) =B_{0}\exp \left(
-\frac{r^{2}}{2d^{2}}\right) $. The two parameters, $d$ and
$B_{0}$, define the half-width and the amplitude of the initial
distribution of the hyperfine field, respectively. Then the
solution of Eq. (\ref{diffusion}) is:

\begin{equation}
B_{hf}\left( r,t\right) =B_{0}e^{-\frac{t}{T_{1}}}\left(
1+\frac{t}{t_{0}} \right) ^{-1}e^{-\frac{r^{2}}{2d^{2}\left(
1+\frac{t}{t_{0}} \right) }}\text{ \ \ ,}  \label{B(r,t)}
\end{equation}
where $t_{0}=\frac{d^{2}}{2D}$. The value of $t_0$ is the time it
takes for the $B_{hf}(0,t)$ to reduce by factor of two from $t=0$
due to the nuclear spin diffusion. The nuclear-spin relaxation
time, $T_{1}$, in semiconductors at sufficiently low temperatures
is rather long. It varies from several hours to a few minutes
\cite{Zahar}. The available experimental values for the diffusion
coefficient are $D \sim 10^{-13}$ cm$^2$s$^{-1}$ for $^{75}$As in
bulk GaAs \cite{Paget82} and $D=10^{-14}$ cm$^2$s$^{-1}$ in
Al$_{0.35}$Ga$_{0.65}$As \cite{malin}. For $d=1$ and 5 $\mu$m
taking $D=10^{-13}$ cm$^2$s$^{-1}$ we have $t_0= 5 \times 10^{4}$,
$1.25 \times 10^{6}$ s.

\section{ENERGY SPECTRUM} \label{sec3}

The microscopic description is based on the following Hamiltonian:

\begin{equation}
H=-\frac{\hbar ^{2}}{2m^{\ast }}\Delta +\frac{1}{2}g^{\ast}\mu
_{B}{\bf \sigma B}_{hf}\left( r,t\right)+U\left( z\right)
\end{equation}
where $m^{\ast }$ is the electron effective mass, $g^{\ast}$ is
the effective electron $g$-factor ($g^{\ast}_{GaAs}=-0.44$),
$\mu_B$ is the Bohr magneton, ${\bf \sigma}$ is the vector of
Pauli matrices, ${\bf B}_{hf}$ is given by Eq. (\ref{B(r,t)}) and
$U(z)$ is the 2DEG confining potential. We suppose, as is usually
done for the 2DEG, that only the lowest sub-band, corresponding to
the confinement in $z$-direction, is occupied and we can ignore
the higher sub-bands. Thus, we omit $z$-dependence of the wave
function in the following. The time scale introduced by a nuclear
spin system is several orders of magnitude larger than the time
scale of typical electron equilibration processes. In such a case
the conduction electrons see a quasi-constant nuclear field. This
simplifies calculation by avoiding the complications which would
appear when solving the Schr\"{o}dinger equation with the time
dependence due to polarized nuclei. We take into account the
electrons of only one spin direction (for which the effective
potential is attractive).

The one-electron eigenvalue problem with the attractive Gaussian
potential (Eq. (\ref{B(r,t)})) does not admit analytical
solutions. Different approximate methods
\cite{Gaus1,Gaus2,Gaus3,Gaus4,Gaus5} were implemented to solve
this problem. In the present paper, an analytical solution of the
Schr\"{o}dinger equation is found within the parabolic
approximation of the hyperfine field \cite{Gaus1}:

\begin{equation}
\widetilde{B}_{hf}=a-br^2 \label{PP}
\end{equation}
connected with Eq. (\ref{B(r,t)}) by the relations:

\begin{equation}
\widetilde{B}_{hf}\left( 0,t \right)=B_{hf}\left( 0,t \right)
\label{depth}
\end{equation}
and

\begin{equation}
\int_0^{r_0}r \widetilde{B}_{hf}\left( 0,t \right)dr=\int_0^\infty
r B_{hf}\left( 0,t \right) dr .\label{width}
\end{equation}
Here $r_0^2=a/b$. Eq. (\ref{depth}) connects the depth of
potentials, Eq. (\ref{width}) provides equal nuclear-spin
polarization for the two fields. From Eqs.
(\ref{depth},\ref{width}) we obtain
$a=B_0\frac{e^{-\frac{t}{T_1}}}{1+\frac{t}{t_0}}$ and
$b=B_0\frac{e^{-\frac{t}{T_1}}}{2d^2\left(1+\frac{t}{t_0}\right)}$.
The energy spectrum for the parabolic potential (\ref{PP}) in
units of $E_0 =\frac{\hbar ^{2}}{2m^{\ast}d^2}$ is given by

\begin{equation}
\varepsilon_{n,m}=-\frac{g^{\ast}\mu
_{B}B_0}{2E_0}\frac{e^{-\frac{t}{T_1}}}{1+\frac{t}{t_0}}+
\sqrt{\frac{g^{\ast}\mu _{B}B_0}{E_0}}
\frac{e^{-\frac{t}{2T_1}}}{1+\frac{t}{t_0}}\left( 2n+ |m|+1
\right) \label{Spectr}
\end{equation}
where $n=0,1,...$ and $m=0,\pm 1,...$ .

The exact solution of the Schr\"{o}dinger equation with the
Gaussian profile of the hyperfine field (Eq. (\ref{B(r,t)})) was
found numerically. Due to the cylindrical symmetry of the problem,
the wave function can be written as $ \psi(\rho
,\phi)=\frac{1}{\sqrt{2\pi}}e^{im \phi} R(\rho)$. The equation for
 the radial part $R(r)$ of wave function has a form

\begin{equation}
\left[ \frac{1}{x}\frac{d}{dx}x\frac{d}{dx}-\frac{m^{2}}{x^{2}}
+\gamma \frac{B_{hf}(x,t)}{B_{hf}(0,0)}+\varepsilon_{n,m} \right]
R_{n,m} =0, \label{The_ODE}
\end{equation}
where $x=r/d$ is the dimensionless coordinate and
$\gamma=\frac{g^{\ast}\mu _{B}B_{hf}(0,0)}{2E_0}$. For $d=1$ and
$5$ $\mu$m, taking $m^*=0.067m_e$, we have $E_0=0.57 \times
10^{-3}$, $0.023 \times 10^{-3}$ meV; for $B_{hf}(0,0)=2.65$ (50\%
nuclear spin polarization) and $5.3$ T (100\% nuclear spin
polarization) corresponding energies are $\frac{1}{2}g^{\ast}\mu_B
B_{hf}(0,0)=3.4 \times 10^{-2}$ and $6.8 \times 10^{-2}$ meV. We
have used the Shooting Method to solve Eq. (\ref{The_ODE}),
subjecting the solution to the following boundary conditions:  $
R_{n,m} \left( \rho\rightarrow 0 \right)=\rho^{|m|}$ and $ R_{n,m}
\left( \rho\rightarrow \infty \right)=0$. The results of the
numerical calculations are presented below.

The time-dependence of the electron energy levels in the NSPIQD is
determined by the time-dependence of the confining hyperfine
field. There are two characteristic times in the problem: the
diffusion characteristic time $t_{0}$ and the relaxation
characteristic time $T_{1}$. We can distinguish the diffusive
regime, when $t \sim t_{0} \ll T_1$, the intermediate regime, $t\
\sim t_0 \sim T_1 $ and the relaxation regime, $t \sim T_{1} \ll
t_{0}$ . Here $t$ is the observation time.

\begin{figure}[tbp]
 \centering
  \includegraphics[width=8cm, angle=270]{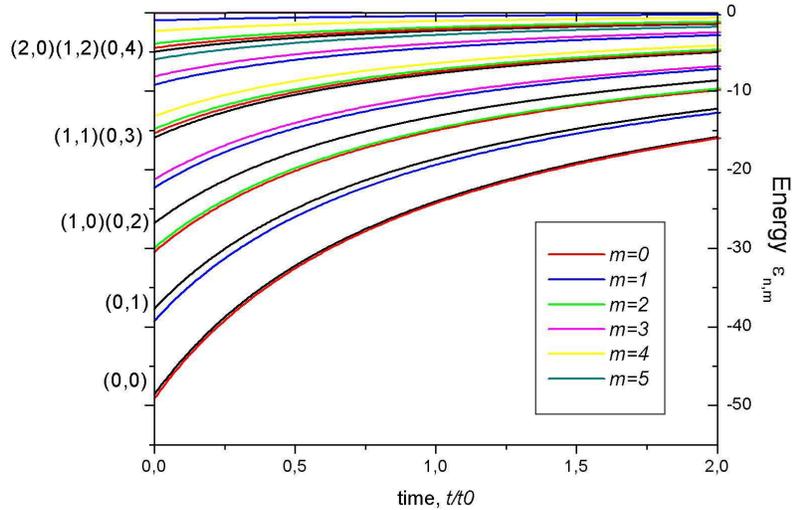}
 \caption{Energy spectra of electrons in NSPIQD with initial half-width $d=1\mu$m
 and $B_{hf}(r=0,t=0)=2.65$T as a function of time in the diffusion regime,
 $T_1/t_0=100$. The black lines are the energy levels for parabolic potential labeled
 by quantum numbers $(n,m)$ at the left. The other lines correspond to the energy
 levels for the Gaussian potential, the lines having a same color have the same quantum
 number $m$; the quantum number $n$ is equal to $0$ for the lowest line of each color and increases
 by $1$ for lines of the same color from bottom to top.}\label{en_diff}
\end{figure}

Figure \ref{en_diff} shows the time dependence of the electron
energy levels for the Gaussian and parabolic potentials in the
diffusion regime. We emphasize that the parabolic potential can be
regarded as a good approximation of the Gaussian potential only
for the ground state. The excited-state energy levels for the
parabolic potential reveal large deviations from those for the
Gaussian potential, which manifest in the degeneracy of states and
in the shift of levels. This result is qualitatively similar to
those obtained for 3D Gaussian and parabolic potential
\cite{Gaus1}. However, time-dependence of energy levels for both
potentials show quite similar behavior. The number of energy
levels in NSPIQD remains constant, whereas their depth decreases.
From Eq. (\ref{Spectr}) it follows that in the diffusion regime
the time-dependence of the energy levels in the parabolic
potential is
$\varepsilon_{n,m}(t)=\frac{\varepsilon_{n,m}(0)}{1+\frac{t}{t_0}}$.
It can be shown that the energy levels in the Gaussian potential
have the same time-dependence. Substituting Eq. (\ref{B(r,t)})
into Eq. (\ref{The_ODE}) and introducing the variable $\xi$ as
$x=\xi \sqrt{1+t/t_0}$ we obtain:

\begin{equation}
\left[
\frac{1}{\xi}\frac{d}{d\xi}\xi\frac{d}{d\xi}-\frac{m^{2}}{\xi^{2}}
+\gamma e^{-\frac{\xi^2}{2}}+\left( 1+\frac{t}{t_0}\right)
\varepsilon_{n,m} \right] R_{n,m} =0. \label{The_ODE1}
\end{equation}
The time-dependent factor, $\left( 1+\frac{t}{t_0} \right)$,
appears in Eq. (\ref{The_ODE1}) only as a product with
$\varepsilon_{n,m}$ thus proving the statement.

\begin{figure}[tbp]
 \centering
 \includegraphics[width=8cm, angle=270]{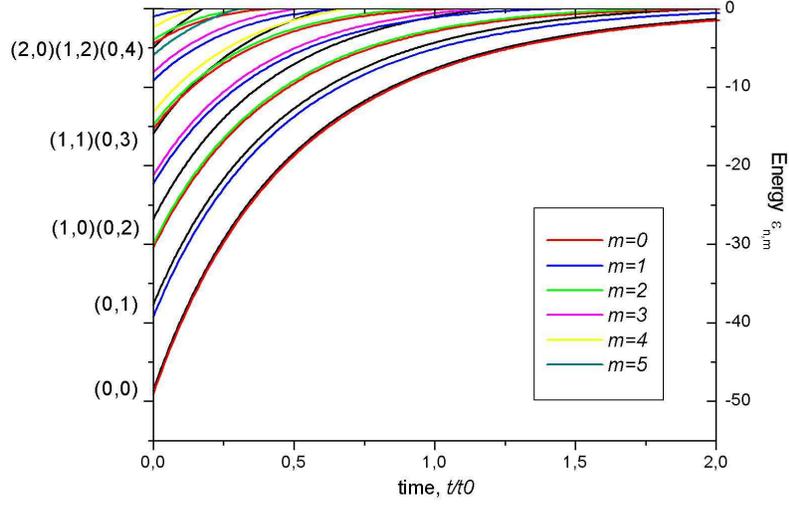}
 \caption{Energy spectra of electrons in NSPIQD as a function of time in the intermediate regime,
 $T_1/t_0=1$. The parameters of calculations and labeling of levels are as on Fig. \ref{en_diff}.}\label{en_interm}
\end{figure}

\begin{figure}[tbp]
 \centering
 \includegraphics[width=8cm, angle=270]{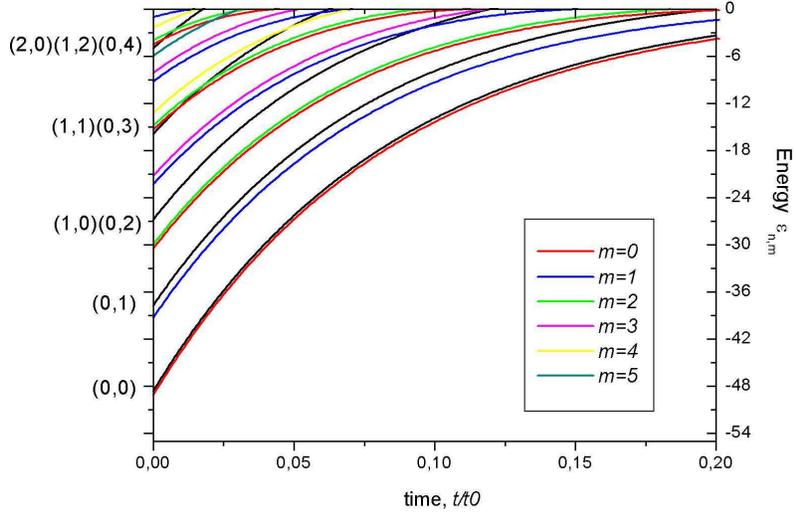}
 \caption{Energy spectra of electrons in NSPIQD as a function of time in the relaxation regime,
 $T_1/t_0=0.1$. The parameters of calculation and labeling of levels are as on Fig. \ref{en_diff}.}\label{en_relax}
\end{figure}

Figures \ref{en_interm} and \ref{en_relax} show the results
obtained for the intermediate and relaxation regimes. On the
contrary, the number of the energy levels in NSPIQD decreases in
time in these regimes. This decrease occurs on the scale of $T_1$.
We can not explicitly obtain time-dependence of energy levels for
the Gaussian potential in these regimes. The parabolic
approximation of the hyperfine field serves as a good
approximation again only for the ground energy level. The
evolution of excited-state energy levels in the Gaussian and in
the parabolic potentials are different: the lifetimes of energy
levels obtained in the case of the parabolic potential are shorter
than in the case of the Gaussian potential.

It is important to know the lifetime of the NSPIQD. We can
consider electron states in the NSPIQD up to the moment when the
confining potential depth is more than the temperature.
Consequently, the lifetime $t_{l}$ of the NSPIQD can be defined by
the following condition: $\frac{\left| g^* \mu_B B_{hf}\left(
0,t_{l}\right)
\right|}{2} =k_{B}T$, where $k_{B}$ is the Boltzman constant and $%
T$ is the temperature. Using Eq.(\ref{B(r,t)}),
we calculate time $t_{l}$ for two limiting cases: $T_{1}\ll t_{0}$ and $%
T_{1}\gg t_{0}$. In the first case (the relaxation regime), $t_{l}
\sim T_{1}\ln \frac{\left|  g^* \mu _{B}B_{0}\right| }{2 k_{B}T}$.
In the second case (the diffusion regime), $%
t_{l}\sim t_0 \left| \frac{ g^* \mu _{B}B_{0}}{2k_{B}T} -1
\right|$. Time-dependence of the half-width of NSPIQD is $d(t)=d
\sqrt{1+t/t_0}$. Let us estimate it at $t=t_l$. For $T^{\ast
}=30mK$ and $B_{0}=2.65$T we have $d(t)=d$ in the relaxation
regime and $d(t)=3.6d$ in the diffusion regime.

\section{CONCLUSIONS} \label{sec4}

We have studied the electron energy levels of a NSPIQD created in
the region of the intersection of a local nuclear spin
polarization with a 2DEG. The properties of the NSPIQD are
time-dependent because of the nuclear spin diffusion and
relaxation. There are two characteristic time and three
corresponding regimes: the diffusion regime, the intermediate
regime and the relaxation regime. In the diffusion regime, the
number of electron energy levels remains constant with time. In
the relaxation and intermediate regime, the number of electron
energy levels decreases with time. Time-dependence of the electron
energy levels in the diffusion regime has a simple form. Since the
characteristic relaxation time is proportional to the square of
the NSPIQD radius at $t=0$, it is possible to create NSPIQDs
operating in different regimes using the same experimental setup.

The numerical estimations allow us to conclude that the system
under study can be realized experimentally. For a hyperfine field
of just a few teslas, the experiment could be made at a
temperature of the order of $10$mK. The modern experimental
technique allows one to create a region with local nuclear spin
polarization of characteristic sizes $\gtrsim 100$nm, making the
NSPIQD having a small size. The spectroscopy of the NSPIQD could
be used to obtain some information about nuclear spin interactions
in solids, for example, the nuclear spin relaxation time and the
nuclear spin diffusion coefficient.

It should be pointed out that a simplified model was used in this
paper to describe the single electron states in the NSPIQD. We
considered the influence of a nuclear spin-related hyperfine field
on the electron states, whereas the electrons could also alter the
nuclear spin dynamics. The well-known examples of such phenomena
are the indirect long-range nuclear-spin interaction,
electron-assisted mechanisms of nuclear spin relaxation and
nuclear spin precession in an effective field created by the
electrons \cite{Slichter}. Investigation of these effects is
beyond the scope of this paper. Results of such investigations
will be published elsewhere.

\section*{ACKNOWLEDGMENTS} \label{sec5}

We acknowledge useful discussions with V. Privman, S. Shevchenko,
and I. Vagner. This research was supported by the National
Security Agency and the Advanced Research and Development Activity
under Army Research Office contracts DAAD-19-99-1-0342 and DAAD
19-02-1-0035, and by the National Science Foundation, grants
DMR-0121146 and ECS-0102500.

\newpage

\end{document}